\documentclass[12pt,preprint]{aastex}

\newcommand\lsim{\lower0.5ex\hbox{$\; \buildrel < \over \sim \;$}}

\shorttitle{Emerging dimmings} \shortauthors{Zhang et al.}

\begin{document}

\title{Emerging dimmings of active regions observed by SDO}

\author{Jun Zhang\altaffilmark{1}, Shuhong Yang\altaffilmark{1},
Yang Liu\altaffilmark{2}, and Xudong Sun\altaffilmark{2}}

\altaffiltext{1}{Key Laboratory of Solar Activity, National
Astronomical Observatories, Chinese Academy of Sciences, Beijing
100012, China; zjun@nao.cas.cn; shuhongyang@nao.cas.cn}

\altaffiltext{2}{W.W. Hansen Experimental Physics Laboratory,
Stanford University, Stanford, CA 94305-4085, USA;
yliu@sun.stanford.edu; xudong@sun.stanford.edu}

\begin{abstract}

With the observations from the Atmospheric Imaging Assembly and the
Helioseismic and Magnetic Imager aboard the \emph{Solar Dynamics
Observatory}, we statistically investigate the emerging dimmings
(EDs) of 24 isolated active regions (IARs) from June 2010 to May
2011. All the IARs present EDs in lower temperature lines (e.g., 171
{\AA}) at their early emerging stages, meanwhile in higher
temperature lines (e.g., 211 {\AA}), the ED regions brighten
continuously. There are two type of EDs: fan-shaped and halo-shaped.
There are 19 fan-shaped EDs and 5 halo-shaped ones. The EDs appear
several to more than ten hours delay to the first emergence of the
IARs. The shortest delay is 3.6 hr and the longest 19.0 hr. The EDs
last from 3.3 hr to 14.2 hr, with a mean duration of 8.3 hr. Before
the appearance of the EDs, the emergence rate of the magnetic flux
of the IARs is from 1.2 $\times$ 10$^{19}$ Mx hr$^{-1}$ to 1.4
$\times$ 10$^{20}$ Mx hr$^{-1}$. The larger the emergence rate is,
the shorter the delay time is. While the dimmings appear, the
magnetic flux of the IARs ranges from 8.8 $\times$ 10$^{19}$ Mx to
1.3 $\times$ 10$^{21}$ Mx. These observations imply that the
reconfiguration of the coronal magnetic fields due to reconnection
between the newly-emerging flux and the surrounding existing fields
results in a new thermal distribution which leads to a dimming for
the cooler channel (171 {\AA}) and brightnening in the warmer
channels.

\end{abstract}


\section{Introduction}

Coronal dimmings have been observed in white light,
extreme-ultraviolet (EUV) lines, and soft X-rays (Hansen et al.
1974; Rust \& Hildner 1976; Sterling \& Hudson 1997; Gopalswamy \&
Hanaoka 1998; Harrison \& Lyons 2000). Filament eruptions and other
ejections of mass from the Sun are often associated with dimmings in
the low corona at many wavelengths (Zarro et al. 1999; Harrison et
al. 2003; Attrill et al. 2006; Reinard \& Biesecker 2009; Dai et al.
2010).

Some dimming regions are co-spatial with sheared magnetic structures
of pre-flare active regions (ARs) and are speculated to be the
remnant signatures of the eruption of large-scale magnetic flux
ropes during coronal mass ejections (CMEs) (Hudson et al. 1998;
Canfield et al. 1999). The evolution timescale of coronal dimmings
is shorter than typical radiative cooling timescale in the corona,
which implies the coronal dimmings are formed due to density
depletion via expansion or ejection rather than temperature decrease
(Hudson et al. 1996). CMEs cause an opening of magnetic field lines
and plasma outflows and subsequent emission decrease. Thus mass loss
is responsible for the dimming process. Harra \& Sterling (2001)
investigated Doppler observations within two coronal dimmings and
found that the dimmings are associated with outflowing material.
Recent results from Jin et al. (2009) also show that, during the CME
eruption, strong outflows are visible in dimming regions at
different heights.

In this Letter, we report the discovery of emerging dimmings (EDs)
of isolated ARs (IARs), with the observations from the Atmospheric
Imaging Assembly (AIA; Lemen et al. 2012) and the Helioseismic and
Magnetic Imager (HMI; Scherrer et al. 2012; Schou et al. 2012)
aboard the {\it Solar Dynamics Observatory} (\emph{SDO}; Pesnell et
al. 2012). The paper is organized as follows. In Section 2 we
describe the observational data we use. The observations are
presented in Section 3. In Section 4, we conclude this study and
discuss the results.

\section{Observations}

SDO/AIA uninterruptedly observes the Sun's full disk at 10
wavelengths with a cadence of 12 s and a pixel size of 0$\arcsec$.6.
The measurements reflect various temperatures of the solar
atmosphere (from $\sim$5000 K to $\sim$2.5 MK) from the photosphere
to the corona. SDO/HMI measures Doppler-velocity, line-of-sight
(LOS) and vector magnetic fields. The data cover the full disk of
the Sun with a spatial sampling of 0$\arcsec$.5 pixel$^{-1}$. The
full disk LOS magnetograms are taken with a cadence of 45 s.

The data adopted here were obtained from June 2010 to May 2011. This
period belongs to the rising phase of the 24th solar cycle, quite a
few of ARs emerged from the solar photosphere. On the other hand,
there were not too many ARs on the solar atmosphere, so we can track
some ARs which emerged isolatedly, termed ``IARs", i.e., no other
pre-existing ARs within 300$\arcsec$ away from the emerging ARs. We
track the emergence of ARs and identify 59 ARs which emerged on the
solar disk during this period. Among them, 24 IARs were detected.

\section{Results}

All the 24 IARs are accompanied by EDs. The parameters about the 24
IARs and the EDs are shown in Table 1. We find that these EDs can be
classified into two types according to their performance during
evolution: fan-shaped EDs (FEDs) and halo-shaped EDs (HEDs). Most
EDs are type one, FEDs. There are 19 FEDs among the 24 EDs, and only
5 EDs are HEDs. For each type, we provide two movies as an example.

\subsection{Type I: FEDs}

One example of FEDs is shown in Figure 1 (see also movies 1 and 3,
available in the online edition). Two AIA 171 images (Figures 1A and
1B) show the coronal response to the emerging AR 11122. This AR
first emerged near N15E16, at 12:00 UT, Nov. 5, 2010. At 18:00 UT,
the AR exhibited as a significant dipolar configuration (Figure 1C).
There were some coronal loops connecting the opposite polarities and
there was no dimming at the emergence region (Figure 1A). After the
emergence progress for 6 hours, an ED appeared. The ED grew and
expanded mainly toward the southeast direction. So it is named FED
due to its fan-shaped structure. The FED was well developed at 01:30
UT, Nov. 6, 2010 (Figure 1B). The HMI LOS magnetograms show that
both the positive and negative magnetic fields are compact, and they
separate alone opposite directions (Figure 1D). The inner boundary
of the FED envelops the emerging AR. The FED lasted 11.3 hr.

When the FED expanded, there were some brightenings (e.g., regions
B1, B2, and B3 outlined in Figure 1B) at the FED front. Figure 1E
shows the light curves of one dimming region (D1) and three
brightening regions (B1, B2 and B3). Beginning at 23:42 UT, the
brightness of D1 decreased from about 0.8 I$_{0}$ to 0.45 I$_{0}$,
where I$_{0}$ represents the average intensity of the quiet Sun. On
the contrary, the brightness of the brightening structures increased
while the dimming appeared.

There was no dimming during the AR emergence in 304 {\AA}, 193 {\AA}
and 211 {\AA} lines that represent the transition region and corona.
Indeed, the average brightness of 304 {\AA} and 193 {\AA} in the
171{\AA} FED region, denoted by the green curve, increases 40\% and
15\%, respectively. The brightness increment in 211 {\AA} images is
more significant, as much as 75\%.

\subsection{Type II: HEDs}

An example of HEDs is displayed in Figure 2 (see movies 2 and 4,
available in the online edition). The HED is associated with AR
11130 which emerged at 04:46 UT, Nov. 27, 2010. At beginning, there
were only several tiny loops at the AR site in the quiet Sun
background (see Figure 2A). About 5 hours later, an ED appeared at
the AR location, and then expanded outward from the AR center. At
14:58 UT, the ED was well developed and appeared as a halo-shaped
structure, the HED surrounding the AR (Figure 2B). The evolution of
underlying magnetic fields are displayed in Figures 2C and 2D. The
magnetograms obtained by the HMI show that both the negative and
positive fields are dispersed, and form a circle shape, different
from AR 11122 shown in Figure 1. There are also some brightening
structures (B1, B2, and B3 in Figure 2B) at the HED front. The light
curves of three brightening regions (B1, B2 and B3) and a dimming
region (D1 outlined in Figure 2B) are displayed in Figure 2E. From
09:00 UT to 12:00 UT, Nov. 27, 2010, the brightness of the D1 region
decreased 35\%, while the bright structures, B1, B2 and B3 exhibited
an increase of brightness about 20\%.

The HED, a dimming area in 171 {\AA} images, appeared as brightening
region in the lines formed in transition region and high corona, as
displayed in Figures 3C, 3E, and 3F. The brightness in 304 {\AA} at
the HED location increased 1.2 times, in 193 {\AA} 1.7 times while
that in 211 {\AA} 3.2 times.

\subsection{Statistical Results of the ERs}

Relationships among the delay time, duration of EDs, the emergence
rate, and magnetic flux are investigated using the sample of those
24 IARs (Figure 4). The delay time is defined to be the time between
the emergence of AR and appearrance of ED. The delay time ranges
from several hours to more than ten hours with the shortest delay
time of 3.6 hr and the longest of 19.0 hr. The EDs last from 3.3 hr
to 14.2 hr, with a mean duration of 8.3 hr. The emergence rate of
the magnetic flux before the appearance of the EDs ranges from 1.2
$\times$ 10$^{19}$ Mx hr$^{-1}$ to 1.4 $\times$ 10$^{20}$ Mx
hr$^{-1}$. When the dimmings appear, the magnetic flux of the IARs
reaches about 8.8 $\times$ 10$^{19}$ Mx to 1.3 $\times$ 10$^{21}$
Mx.

There is a negative relationship with a correlation coefficient
$\alpha$ of $-0.53$ between the delay time and the emergence rate of
the IARs (Figure 4A). In other words, the larger the emergence rate
is, the shorter the delay time is. Figure 4C shows a similar
relationship between delay time and magnetic flux, and the
correlation coefficient $\alpha$ is $-0.48$. On the contrary, there
exists a positive relationship between the duration of the EDs and
the magnetic flux of the IARs (Figure 4D), and the correlation
coefficient $\alpha$ is $0.60$. It means that an ED lasts a longer
time, if the magnetic flux at the ED appearance is larger. As shown
in Figure 4B, the relationship between the duration of the EDs and
the emergence rate is very weak.

\section{Conclusions and Discussion}

We statistically investigate the EDs associated with IARs from June
2010 to May 2011, and 24 IARs have been identified during the
one-year period. All the IARs present the EDs at their early
emerging stages. There are two type of EDs: fan-shaped and
halo-shaped. Usually the EDs appear several to more than ten hours
delay to the first emergence of the IARs. The EDs last from 3.3 hr
to 14.2 hr, with a mean duration of 8.3 hr. The HMI observations
show that, the larger the emergence rate is, the shorter the delay
time is. While the dimmings appear, the magnetic flux of the IARs
ranges from 8.8 $\times$ 10$^{19}$ Mx to 1.3 $\times$ 10$^{21}$ Mx.

There are several differences between EDs and ``classical dimmings".
First, EDs are only relevant to the emergence of ARs, and no
filament eruptions, no flares and no CMEs are accompanying. The
``classical dimmings" usually accompany with eruptive events, such
as filament eruptions, flares, and CMEs (Jiang et al. 2007; Attrill
et al. 2007). In this work, EDs appear with the emergence of AR.
Second, EDs are much more pronounced in 171 {\AA} than in the
higher-temperature lines, 193 {\AA} and 211 {\AA} (see Figure 3).
More precisely, at the ED regions, the high temperature line
observations show that a heating process happens at high coronal
level, as the brightness increases when the EDs appear. This
suggests that, at the early stage of AR emergence, the density of
lower temperature plasma decreases, leading to the appearance of
EDs. On the other hand, the plasma temperature at high corona
increases as revealed by the brightening in 193 {\AA} and 211 {\AA}
images. However, according to previous studies, the ``classical
dimmings" are much more pronounced in 195 {\AA} than in the
lower-temperature line 171 {\AA} (Robbrecht \& Wang 2010). We
conclude that most of the hot coronal plasma is not ejected.

Besides the differences, some similar behaviors between EDs and
``classical dimmings" remain. First, the lifetime of both EDs and
``classical dimmings" ranges from several hours to more than ten
hours (Thompson et al. 2000; Reinard \& Biesecker 2008). With a
sample of 96 CME-associated EUV coronal dimmings between 1998 and
2000, Reinard \& Biesecker (2008) reported the durations of the
dimmings typically range from 3 to 12 hr, which is consistent with
the range of 3.3 to 14.2 hours for EDs. Second, at the outermost
edge of ED and the ``classical dimming" regions, there are EUV
brightenings (Attrill et al. 2007; Jiang et al. 2007; Cohen et al.,
2009), which may be signatures of magnetic reconnection.

As shown in this study, the EDs can be classified into two types,
i.e., FEDs and HEDs. This is probably due to the different
configurations of their source regions. All the 19 IARs associated
with FEDs emerged as dipolar regions with compact opposite
polarities (e.g., AR 11122 in Figure 1). The 5 IARs associated with
HEDs have a circle-like configuration (e.g., AR 11130 in Figure 2).
When an AR emerges, the magnetic fields of emerging ARs reconnects
with the surrounding existing fields. So a new thermal distribution
is created, as suggested by Schrijver et al. (2010). The structures
in hot and cool wavelength images represent plasma at different
temperature. The magnetic reconnection between the emerging and
background fields heats up the coronal plasma, so that the
cool-wavelength 171 {\AA} appears dimming (without enough cool
plasma) while the hot-wavelengths brightens up (most plasma has been
heated up). This process leads to a dimming for the cooler channel
and brightnening in the warmer channels. In 171 {\AA} wavelength,
there are some structures in the dimming regions, i.e., in Figure 1B
the main dimming region displays a loop-shape appearance. While in
193 and 211 {\AA} wavelengths (see movie 3), there are many loop
systems. By comparison, we can see that the 171 {\AA} dimming
structures are consistent with the 193 and 211 {\AA} bright
structures. Bright structures are usually considered as magnetic
loop systems in the solar corona. So the dimming areas are occupied
by magnetic loop systems.

In order to comprehensively understand this phenomenon, some
spectral observations (such as Hinode/EIS) are required to diagnose
the temperature, density, and flow velocity in the emerging
dimmings. Moreover, studying the variation of the extrapolated
coronal magnetic field configurations during the dimming process by
using vector magnetic fields is also under consideration in future
work.

\acknowledgments {We are grateful to the referee for useful
suggestions. We thank Drs. Carolus J. Schrijver and Jiong Qiu for
helpful comments. This work is supported by the National Basic
Research Program of China under grant 2011CB811403, the National
Natural Science Foundations of China (11025315, 40890161, 10921303,
41074123, 11203037 and 11003024), and the CAS Project KJCX2-EW-T07.
The data are used by courtesy of NASA/SDO and the AIA and HMI
science teams.}

{}

\clearpage

\begin{table}
\begin{center}
\caption{Parameters about the 24 IARs and the EDs.\label{tbl-1}}
\vspace{0.2 cm} \centering
\begin{tabular}{ccccccc}
\tableline\tableline
IARs & Position & FET\tablenotemark{a} & EDT\tablenotemark{b} & EDD\tablenotemark{c} & FER\tablenotemark{d} & ED type\\
     & (at FET) & (UT) & (UT) & (hr) & (10$^{19}$ Mx hr$^{-1}$) & \\
\tableline
AR11079 & S26W06 & 2010-06-07 12:00 & 2010-06-07 19:48 & 3.3 & 4.0 & I \\
AR11081 & N25W38 & 2010-06-11 00:30 & 2010-06-11 05:24 & 7.3 & 14.0 & I \\
AR11086 & N18W30 & 2010-07-03 22:00 & 2010-07-04 06:42 & 7.1 & 4.0 & I \\
AR11088 & N18E46 & 2010-07-11 01:24 & 2010-07-11 07:06 & 12.2 & 6.3 & II \\
AR11105 & N18E02 & 2010-09-01 12:18 & 2010-09-02 01:06 & 9.1 & 2.5 & I \\
AR11114 & S22W25 & 2010-10-13 08:00 & 2010-10-14 03:00 & 8.6 & 1.3 & I \\
AR11116 & N22E05 & 2010-10-16 10:00 & 2010-10-16 16:06 & 6.4 & 4.2 & I \\
AR11122 & N15E16 & 2010-11-05 12:00 & 2010-11-05 23:42 & 11.3 & 2.4 & I \\
AR11130 & N13E20 & 2010-11-27 04:46 & 2010-11-27 09:43 & 10.4 & 4.5 & II \\
AR11132 & N11E15 & 2010-12-03 13:42 & 2010-12-03 18:06 & 12.3 & 4.5 & II \\
AR11142 & S14E60 & 2010-12-29 21:00 & 2010-12-30 02:00 & 7.4 & 6.6 & I \\
AR11143 & S22E50 & 2011-01-05 14:18 & 2011-01-05 18:18 & 9.6 & 8.9 & I \\
AR11144 & S16W05 & 2011-01-07 15:24 & 2011-01-07 19:06 & 7.2 & 9.5 & II \\
AR11148 & S28W12 & 2011-01-16 10:00 & 2011-01-16 19:24 & 10.8 & 2.9 & I \\
AR11153 & N15E12 & 2011-02-02 19:42 & 2011-02-03 00:42 & 9.1 & 5.2 & I \\
AR11179 & N09E30 & 2011-03-20 20:00 & 2011-03-21 07:48 & 5.4 & 3.7 & I \\
AR11194 & S32W10 & 2011-04-12 22:00 & 2011-04-13 02:06 & 5.3 & 2.7 & I \\
AR11198 & S26W30 & 2011-04-21 06:30 & 2011-04-21 10:06 & 8.9 & 5.5 & II \\
AR11202 & N15E12 & 2011-04-25 05:42 & 2011-04-25 12:48 & 7.4 & 1.2 & I \\
AR11211 & S12E05 & 2011-05-07 22:00 & 2011-05-08 02:42 & 6.4 & 2.5 & I \\
AR11214 & S20E20 & 2011-05-13 12:42 & 2011-05-13 16:42 & 14.2 & 9.2 & I \\
AR11215 & S23E07 & 2011-05-12 00:00 & 2011-05-12 08:06 & 6.7 & 1.5 & I \\
AR11220 & N14W45 & 2011-05-21 18:00 & 2011-05-22 01:42 & 6.7 & 5.1 & I \\
AR11221 & S18W36 & 2011-05-21 19:00 & 2011-05-22 05:06 & 6.0 & 3.2 & I \\


 \tableline
\end{tabular}
\tablenotetext{a}{FET $=$ Flux first Emerging Time.}
\tablenotetext{b}{EDT $=$ Emerging Dimming first appearance Time.}
\tablenotetext{c}{EDD $=$ Emerging Dimming Duration.}
\tablenotetext{d}{FER $=$ Flux Emerging Rate.}
\end{center}
\end{table}

\clearpage

\begin{figure}
\epsscale{0.8} \plotone{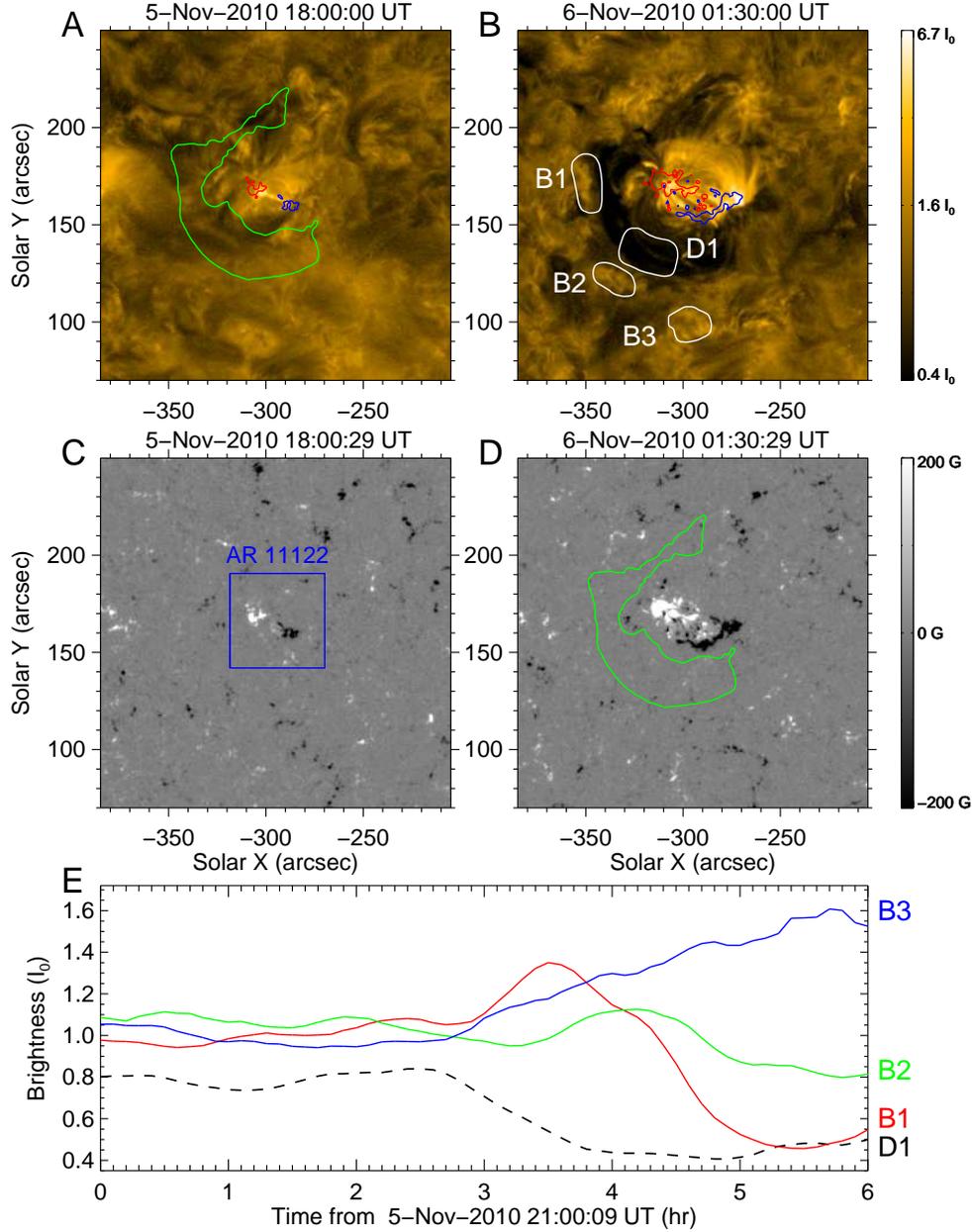}\caption{Example of an FED. Panels
(A) and (B): Two AIA 171 {\AA} images showing the coronal response
to the emerging active region AR 11122. The red and blue contours
represent the positive and negative magnetic fields. The green curve
outlines the FED at 01:30 UT, Nov. 6, 2010. The four white curves
outline a dimming region (D1) and three brightening regions (B1, B2
and B3) where the light curves are measured. The unit I$_{0}$
represents the average intensity of the quiet Sun. Panels (C) and
(D): Corresponding line-of-sight magnetograms obtained by the
\emph{SDO}$/$HMI. The blue window delineates the IAR area. Panel
(E): Light curves of the four given regions in panel (B).
\label{fig1}}
\end{figure}

\clearpage

\begin{figure}
\epsscale{0.8} \plotone{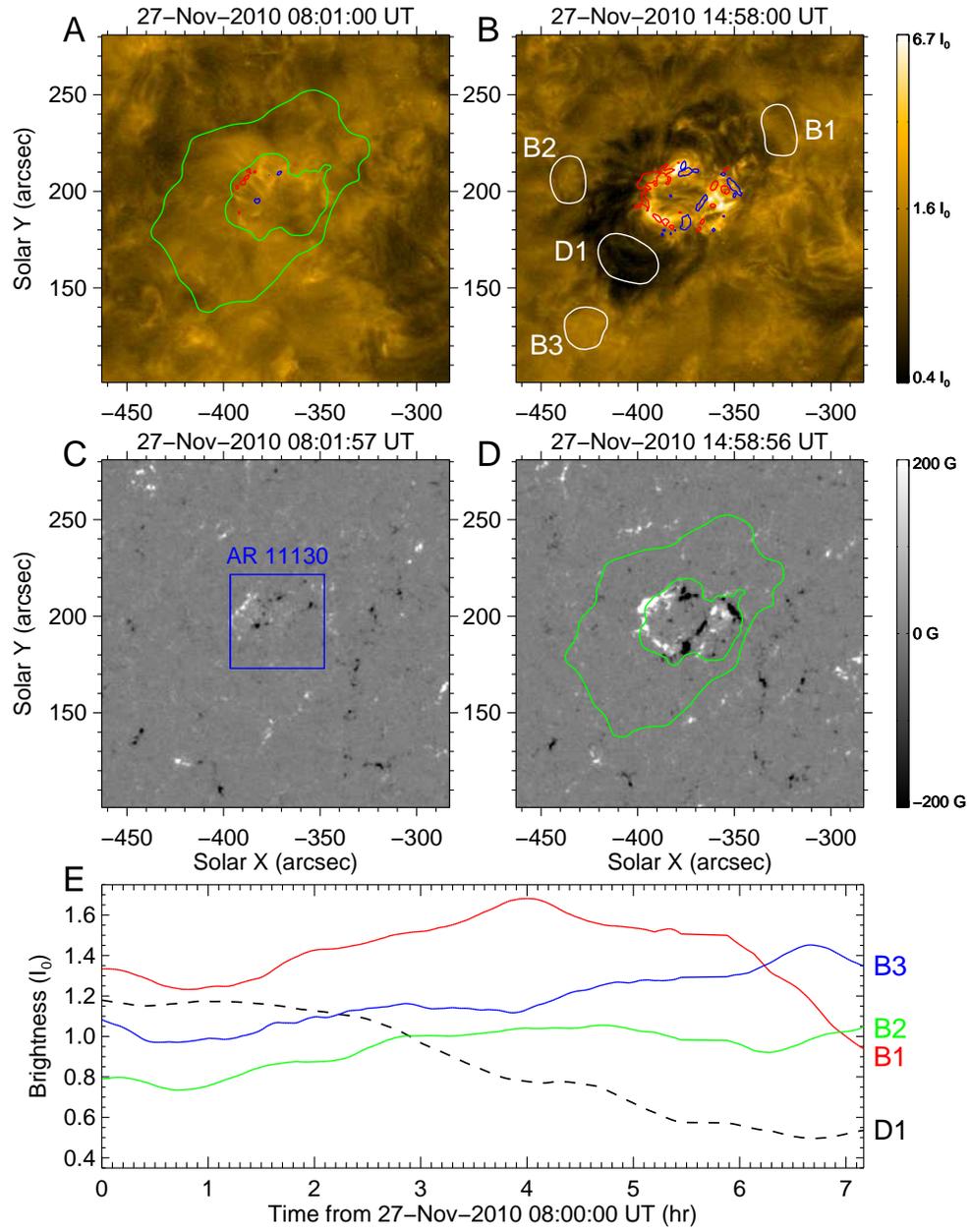}\caption{ Similar to Figure 1 but
for an HED on Nov. 27, 2010. \label{fig2}}
\end{figure}

\clearpage

\begin{figure}
\centering
\includegraphics
[bb=77 176 512 610,clip,angle=0,scale=0.95]{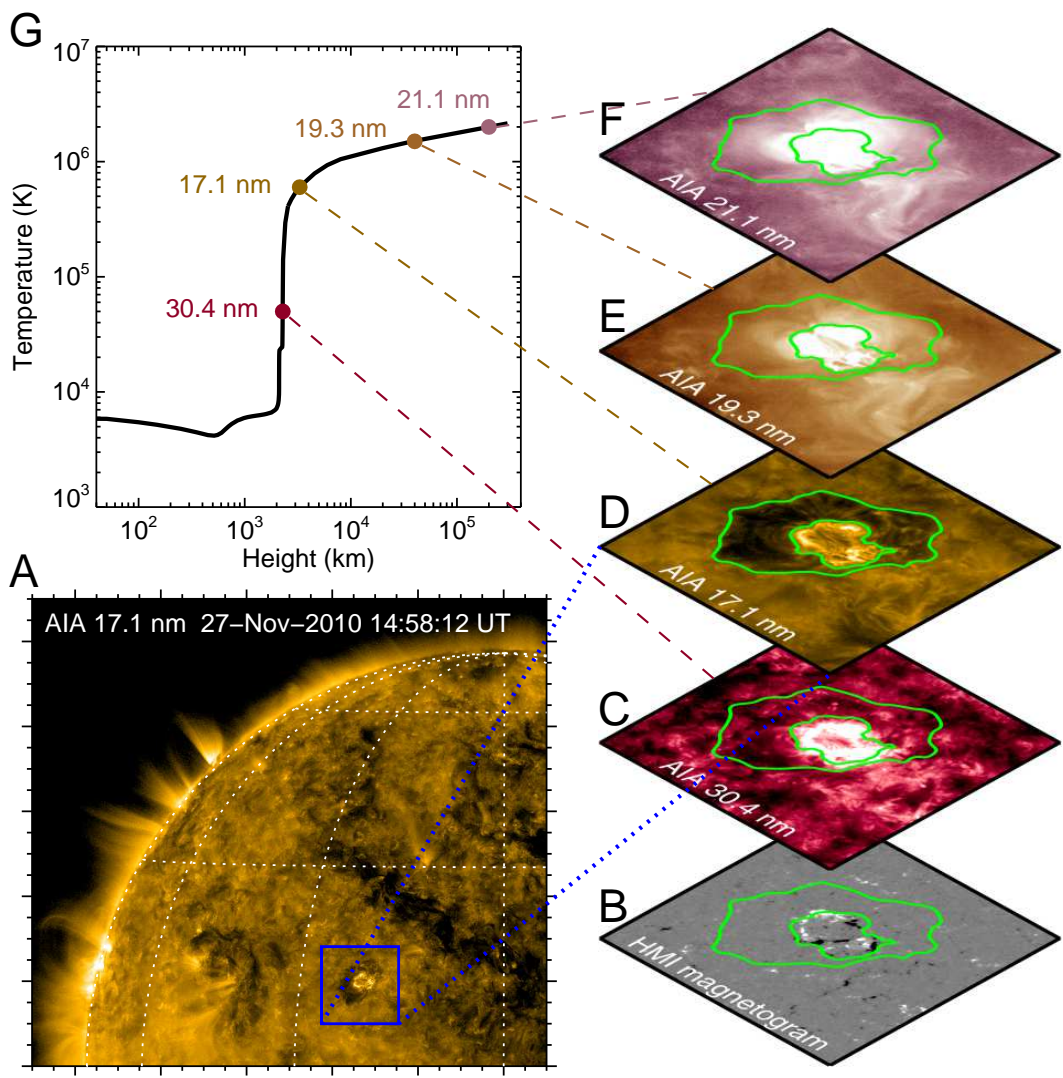}
\begin{flushleft}
\caption{Coronal response to AR 11130 (see Figure 2) in different
temperature lines while ED is well developed. Panel (A): Part of the
full-disk AIA 171 {\AA} image. The blue window outlines the emerging
AR 11130. Panel (B): Corresponding HMI magnetogram of AR 11130.
Panels (C) to (F): Layered atmosphere from the transition region
(panel (C) 304 {\AA}), through the low corona (panel (D) 171 {\AA}),
to the high corona (panel (E) 193 {\AA} and panel (F) 211 {\AA}).
Panel (G): Variation of the temperature versus the height above the
$\tau_{5000}$=1 (optical depth unity in the continuum at 5000 {\AA})
surface. The peak formation temperatures of the 304 {\AA} line
emitted in the transition region, and 171 {\AA}, 193 {\AA}, 211
{\AA} lines which are all emitted in the corona, are 0.6 MK, 1.5 MK,
and 2.0 MK, respectively. \label{fig3}}
\end{flushleft}
\end{figure}

\label{fig3}

\clearpage

\begin{figure}
\epsscale{1.0} \plotone{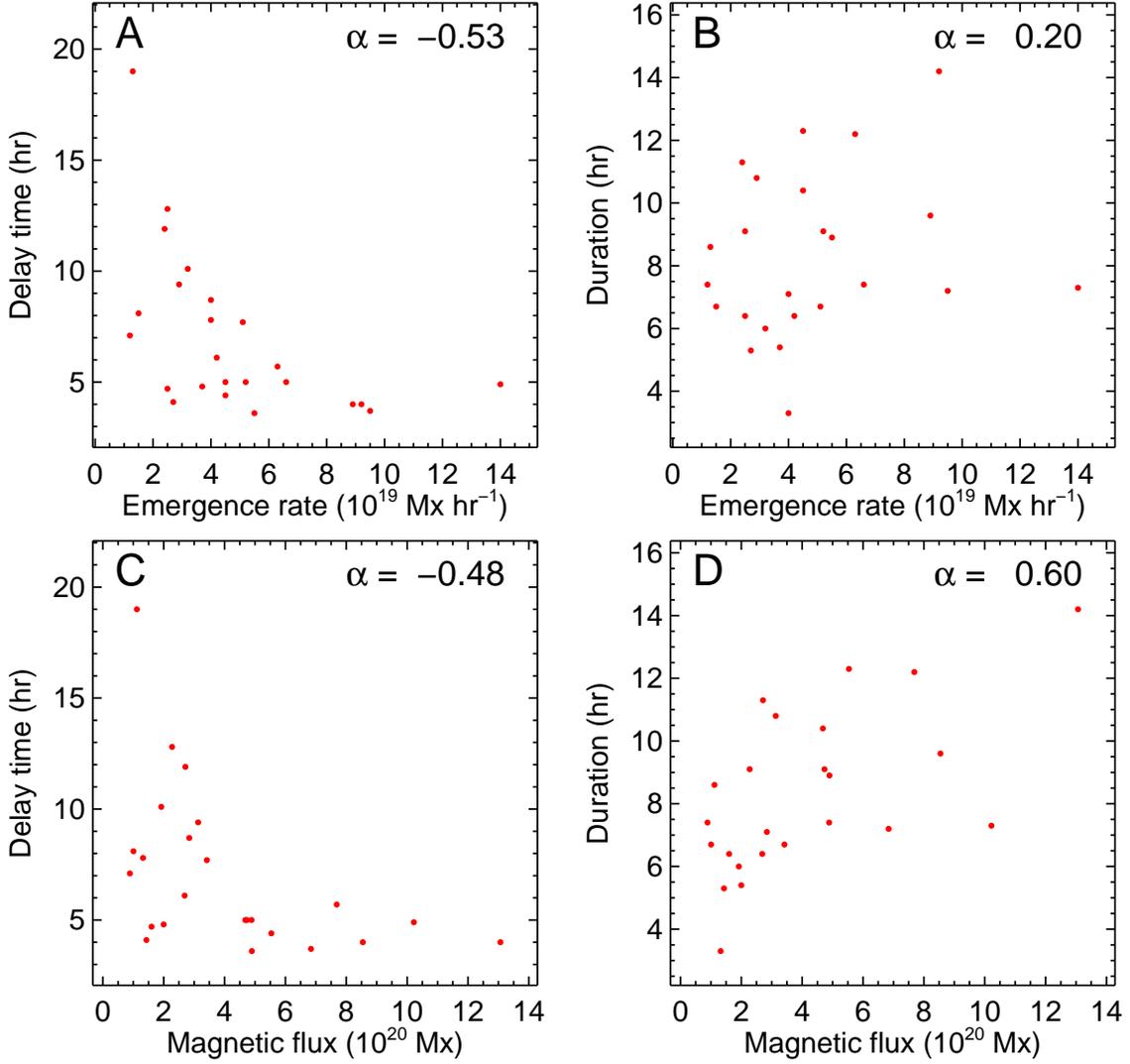} \caption{Panel (A): Relationship
between the delay time of EDs and the magnetic flux emergence rate
of 24 IARs. There is a negative relationship with a correlation
coefficient $\alpha$ of $-0.53$. Panel (B): Relationship between the
duration of EDs and the emergence rate, with a correlation
coefficient $\alpha$ of $0.20$. Panel (C): Relationship between the
delay time of EDs and the magnetic flux of 24 IARs. The correlation
coefficient $\alpha$ is $-0.48$. Panel (D): Relationship between the
duration of EDs and the magnetic flux of 24 IARs. There is a
positive relationship with a correlation coefficient $\alpha$ of
$0.60$. \label{fig4}}
\end{figure}

\clearpage

\end{document}